\documentclass[nofootinbib, reprint, preprintnumbers, amsmath, amssymb, amsfonts, aps, pra, superscriptaddress, floatfix]{revtex4-2}
\usepackage[utf8]{inputenc}
\usepackage{mathtools}
\usepackage{graphicx}
\usepackage{dcolumn}
\usepackage{bm}
\usepackage{physics}
\usepackage{color}
\usepackage{xcolor}
\usepackage{braket}
\usepackage[normalem]{ulem}
\usepackage{enumerate}
\usepackage{enumitem}
\usepackage[colorlinks=true,linkcolor=blue,citecolor=blue,urlcolor =blue]{hyperref}
\usepackage{mathrsfs}
\usepackage{bbm}
\usepackage{siunitx}
\usepackage{listings}
\usepackage{ulem}
\usepackage{tcolorbox}

\usepackage{comment}
\usepackage{lipsum}

\usepackage{placeins}
\begin{document}
\title{Improving the Dynamic Range of Interacting Quantum Sensors via Optimized Quantum Control}

\author{Yasuhiro Aota}
\affiliation{Department of Electrical, Electronic, and Communication Engineering, Graduate School of Science and Engineering, Chuo University, 1-13-27, Kasuga, Bunkyo-ku, Tokyo 112-8551, Japan}

\author{Yuichiro Mori}
\affiliation{Department of Electrical, Electronic, and Communication Engineering, Faculty of Science and Engineering, Chuo University, 1-13-27, Kasuga, Bunkyo-ku, Tokyo 112-8551, Japan}

\author{Takahiko Satoh}
\affiliation{Faculty of Science and Technology, Keio University, Yokohama, Kanagawa 223-8522, Japan}

\author{Yuichiro Matsuzaki}
\email{ymatsuzaki872@g.chuo-u.ac.jp}
\affiliation{Department of Electrical, Electronic, and Communication Engineering, Faculty of Science and Engineering, Chuo University, 1-13-27, Kasuga, Bunkyo-ku, Tokyo 112-8551, Japan}

\author{Hideaki Kawaguchi}
\email{hikawaguchi@g.ecc.u-tokyo.ac.jp}
\affiliation{Medical Research Center for Pre-Disease State (Mebyo) AI, Graduate School of Medicine, The University of Tokyo, 7-3-1, Hongo, Bunkyo-ku, Tokyo 113-0033, Japan}
\affiliation{Graduate School of Science and Technology, Keio University, Yokohama, Kanagawa 223-8522, Japan}

\date{\today}

\begin{abstract}
Interacting spin ensembles are promising platforms for quantum sensing, but spin–spin interactions can induce complex, nonmonotonic responses to external fields, limiting the dynamic range over which the fields can be estimated unambiguously. Here, we propose a quantum-circuit-learning-inspired method to improve the dynamic range of interacting quantum sensors using optimized global control fields. We consider an ensemble of dipole-coupled spins subjected to an inhomogeneous magnetic field generated by an electric current. By optimizing the global control fields applied during sensing, we train the expectation value of the collective magnetization to approximate a prescribed monotonic target function. We also introduce a penalty on the magnetization variance to account for measurement fluctuations. Numerical simulations demonstrate that the optimized control can substantially enlarge the dynamic range compared with sensing without control. We further investigate the trade-off between dynamic range and estimation uncertainty and show that the variance penalty can reduce the uncertainty at the expense of dynamic range. These results demonstrate that learning-assisted global control provides a promising approach to improving the performance of interacting quantum sensors.
\end{abstract}

\maketitle

\section{Introduction}
\label{sec:intro}

Quantum metrology is an essential application of quantum information processing~\cite{giovannetti2011advances,taylor2016quantum,simon2017quantum}.
In a typical quantum sensing protocol, an external field imprints a phase on a quantum probe, and the field strength is inferred from measurement outcomes~\cite{degen2017quantum}.
Spin-based quantum sensors are particularly attractive for detecting weak magnetic fields with high spatial resolution~\cite{taylor2008high,maze2008nanoscale,balasubramanian2008nanoscale}.

Improving sensitivity by increasing the number of sensing spins while retaining spatial resolution motivates the use of dense spin ensembles~\cite{barry2020sensitivity,acosta2009diamonds,mitchell2020colloquium}.
Entanglement-enhanced spin ensembles have also been investigated for detecting individual target spins through their spatially inhomogeneous magnetic fields~\cite{hakoshima2020single,hakoshima2020efficient,hakoshima2021proposed}.
At high densities, however, spin--spin interactions become increasingly important and complicate the sensor dynamics~\cite{bauch2020decoherence,hayashi2020experimental,yoshinaga2022emergence,spielman2024quantum}.
In an interacting sensor, the expectation value of a measured observable can vary nonmonotonically with the field, making different input values difficult to distinguish and limiting the dynamic range (DR) over which the input can be estimated unambiguously~\cite{sugiyama2015precision}.
Coherent-control and Hamiltonian-engineering techniques have been developed to improve sensing with strongly interacting spin ensembles~\cite{zhou2020quantum,biteri2024microscale,zhou2023robust,lu2025dynamical}.
In parallel, quantum phase-estimation and adaptive sensing protocols have addressed the trade-off between sensitivity and dynamic range~\cite{ma2025adaptive,said2011nanoscale,waldherr2012high}.
Quantum deamplification has also been proposed to extend the dynamic range while limiting the associated sensitivity cost~\cite{liu2025deamplification}.

Quantum circuit learning (QCL) provides a framework in which the parameters of a quantum model are trained to approximate a prescribed function of an input~\cite{mitarai2018quantum}.
QCL belongs to the broader field of quantum machine learning (QML), which includes circuit-based, variational, and kernel-based approaches~\cite{schuld2015introduction,schuld2019quantum,carleo2019machine,perez2020data,biamonte2017quantum,cerezo2021variational,havlivcek2019supervised,schuld2021supervised,jerbi2023quantum}.
Previous studies have explored the learning capabilities of quantum models and learning from quantum experimental data~\cite{liu2021rigorous,jager2023universal,huang2022quantum}, as well as challenges involving barren plateaus, quantum data, and kernel methods~\cite{mcclean2018barren,cerezo2021cost,larocca2025barren,huang2021power,thanasilp2024exponential,suzuki2024quantum}.
Learning and variational optimization have also been applied to quantum metrology, including the optimization of sensing states, measurement protocols, and estimation performance~\cite{kaubruegger2019variational,koczor2020variational,ma2021adaptive,kaubruegger2021quantum,maclellan2024endtoend}.
In particular, variational control has been studied in dipolar-interacting spin ensembles~\cite{zheng2022preparation,srivastava2025variational}, and the estimation range has been incorporated into metrological optimization~\cite{kaubruegger2021quantum}.
Other learning-based approaches have addressed broad-range phase estimation~\cite{yang2026global} and task-specific quantum sensing~\cite{khan2026computational}.
These advances motivate a complementary question: can parameterized global control reshape the input--output response of a dense, interacting spin ensemble under fixed collective readout to extend its dynamic range?

Here, we investigate a QCL-inspired response-shaping method for an interacting spin sensor.
We consider spins coupled by magnetic dipole--dipole interactions and subjected to an inhomogeneous magnetic field generated by an electric current $I$.
A parameterized global control field acts during the sensing interval, while both the input field and the spin--spin interactions remain present.
At the end of sensing, the collective $z$ magnetization is measured, and its expectation value defines the current-dependent response.
Using known currents, we optimize the control parameters so that this response approaches a prescribed monotonic target function.
We also consider a penalty on the magnetization variance to account for measurement fluctuations.
The optimized parameters are subsequently held fixed when estimating an unknown current.
While the control parameters are optimized to reproduce the target function, the actual sensing performance is independently evaluated in terms of the DR and estimation uncertainty.

Numerical simulations of small interacting spin ensembles show that optimized global control can reshape the magnetization response toward the target function and enlarge the DR under the studied conditions.
By varying the variance penalty and comparing with uncontrolled sensing at shorter sensing times, we characterize the trade-off between DR and per-measurement estimation uncertainty.
These results demonstrate the feasibility of learning-inspired response shaping in the specified model.

\section{Quantum metrology}
\label{sec:quantum_metrology}

\subsection{Single-spin Ramsey sensing}
\label{subsec:single_spin_ramsey}

We first review the basic principle of Ramsey sensing using a single spin-$1/2$ particle~\cite{degen2017quantum}. Consider a spin subjected to a magnetic field generated by an electric current $I$. We assume that the field is proportional to $I$ and directed along the $z$ axis for positive current. In angular-frequency units with $\hbar=1$, the Hamiltonian is
\begin{equation}
 \hat H_{\mathrm R}^{(1)}(I)
 =\frac{hI}{2}\hat\sigma_z,
 \label{eq:ramsey_single_hamiltonian}
\end{equation}
where $h>0$ is the current-to-angular-frequency coupling strength and $\hat\sigma_z$ is the Pauli-$z$ operator.

We initialize the spin along the $x$ axis, perpendicular to the magnetic field, in the state $|+x\rangle=(|0\rangle+|1\rangle)/\sqrt{2}$. It then precesses about the $z$ axis for a sensing time $\tau$. The resulting state is
\begin{equation}
 |\psi(I,\tau)\rangle
 =\exp[-i\hat H_{\mathrm R}^{(1)}(I)\tau]|+x\rangle.
\end{equation}

We read out the spin component along the $y$ axis by measuring the Pauli-$y$ operator, obtaining the current-dependent expectation value
\begin{equation}
 f_{\mathrm R}^{(1)}(I)
 =\langle\psi(I,\tau)|\hat\sigma_y|\psi(I,\tau)\rangle
 =\sin(hI\tau).
 \label{eq:ramsey_single_response}
\end{equation}
The response is monotonic around $I=0$, specifically for $|hI\tau|<\pi/2$. Within a known interval where the response is one-to-one, the current can be inferred unambiguously from its expectation value. Outside such an interval, different current values may produce the same expectation value. This ambiguity illustrates why the monotonicity of the sensor response is important for its dynamic range (DR).
For example, if $\overline m$ is the sample mean of repeated Pauli-$y$ measurement outcomes, the current can be estimated within this monotonic interval as $\hat I=\arcsin(\overline m)/(h\tau)$~\cite{sugiyama2015precision}.

We next consider how repeated measurements affect the local estimation uncertainty.
Let $T_{\mathrm{total}}$ be the total experimental time and $t_{\mathrm{cycle}}$ the duration of one measurement cycle. The number of repetitions is then approximately $N=T_{\mathrm{total}}/t_{\mathrm{cycle}}$. 
Assuming $N$ independent measurements of the observable $\hat M$ under identical sensing conditions, we define its variance as $\operatorname{Var}_I(\hat M)=\langle\hat M^2\rangle_I-\langle\hat M\rangle_I^2$, where the subscript $I$ indicates that the expectation values are evaluated in the quantum state after sensing with current $I$. The variance of the sample mean is then $\operatorname{Var}_I(\hat M)/N$.
Using error propagation, the local estimation uncertainty is

\begin{equation}
 \delta I_N(I)
 =\frac{\sqrt{\operatorname{Var}_I(\hat M)}}
 {\sqrt{N}\,|\partial_I f(I)|},
 \label{eq:metrology_local_uncertainty}
\end{equation}
where $f(I)=\langle\hat M\rangle_I$. For the single-spin Ramsey measurement, $\hat\sigma_y^2=\mathbb I$ gives
\begin{equation}
 \operatorname{Var}_I(\hat\sigma_y)
 =1-[f_{\mathrm R}^{(1)}(I)]^2
 =\cos^2(hI\tau).
\end{equation}
Using the derivative of Eq.~\eqref{eq:ramsey_single_response}, we obtain
\begin{equation}
 \delta I_{\mathrm R,N}^{(1)}(I)
 =\frac{\sqrt{\operatorname{Var}_I(\hat\sigma_y)}}
 {\sqrt{N}\,|\partial_I f_{\mathrm R}^{(1)}(I)|}
 =\frac{1}{h\tau\sqrt{N}}.
 \label{eq:ramsey_single_uncertainty}
\end{equation}
This expression holds where the response slope is nonzero.

\subsection{Role of preparation and readout axes}
\label{subsec:single_spin_cosine}

The form of the sensing response depends on the directions of the magnetic field, initial spin polarization, and measured observable. Consider a single spin initialized along the $z$ axis in the state $|0\rangle=|+z\rangle$ and subjected to a current-dependent magnetic field along the $x$ axis. The Hamiltonian is
\begin{equation}
 \hat H_{\mathrm C}^{(1)}(I)
 =\frac{hI}{2}\hat\sigma_x,
 \label{eq:cosine_single_hamiltonian}
\end{equation}
where $h>0$ is the current-to-angular-frequency coupling strength. After a sensing time $\tau$, we measure the Pauli-$z$ operator. The expectation value is
\begin{equation}
 f_{\mathrm C}^{(1)}(I)
 =\langle 0|
 e^{i\hat H_{\mathrm C}^{(1)}(I)\tau}
 \hat\sigma_z
 e^{-i\hat H_{\mathrm C}^{(1)}(I)\tau}
 |0\rangle
 =\cos(hI\tau).
 \label{eq:single_spin_cosine}
\end{equation}

Unlike the sinusoidal response in Eq.~\eqref{eq:ramsey_single_response}, this response has a turning point at $I=0$, where its derivative vanishes.

For this measurement, $\hat\sigma_z^2=\mathbb I$ gives
\begin{equation}
 \operatorname{Var}_I(\hat\sigma_z)
 =1-[f_{\mathrm C}^{(1)}(I)]^2
 =\sin^2(hI\tau).
\end{equation}
The local estimation uncertainty is therefore
\begin{equation}
 \delta I_{\mathrm C,N}^{(1)}(I)
 =\frac{\sqrt{\operatorname{Var}_I(\hat\sigma_z)}}
 {\sqrt{N}\,|\partial_I f_{\mathrm C}^{(1)}(I)|}
 =\frac{1}{h\tau\sqrt{N}},
 \label{eq:cosine_single_uncertainty}
\end{equation}
at points where the response slope is nonzero.

Since both the response and the local estimation uncertainty are symmetric under $I\to-I$, the positive- and negative-current monotonic branches are equivalent when evaluated using the same criteria based on response monotonicity and local estimation uncertainty.

\subsection{Noninteracting Ramsey spin ensembles}
\label{subsec:noninteracting_ramsey}

We now extend the single-spin Ramsey protocol introduced in Sec.~\ref{subsec:single_spin_ramsey} to an ensemble of $L$ noninteracting spins. For simplicity, we first neglect the spatial extent of the ensemble and assume that all spins experience the same magnetic field. Using a common coordinate system with the $z$ axis along the field, the Hamiltonian becomes
\begin{equation}
 \hat H_{\mathrm R}(I)
 =\frac{hI}{2}\sum_{j=1}^{L}\hat\sigma_z^{(j)},
 \label{eq:ramsey_multi_hamiltonian}
\end{equation}
where $\hat\sigma_z^{(j)}$ acts on spin $j$.
All spins are prepared along $x$ in $|+x\rangle$, precess about the common $z$ axis, and are read out along $y$ through the collective observable
\begin{equation}
 \hat M_{\mathrm R}
 =\sum_{j=1}^{L}\hat\sigma_y^{(j)}.
\end{equation}

Because all spins undergo the same evolution, each produces the same expectation value $\sin(hI\tau)$ for its Pauli-$y$ measurement. The collective expectation value is therefore the sum of these $L$ identical contributions, giving
\begin{equation}
 f_{\mathrm R}(I)
 =\langle\hat M_{\mathrm R}\rangle_I
 =L\sin(hI\tau).
 \label{eq:ramsey_uniform_response}
\end{equation}

Since the spins are independent, the variance of the collective observable is the sum of the individual variances:
\begin{equation}
 \operatorname{Var}_I(\hat M_{\mathrm R})
 =\sum_{j=1}^{L}\operatorname{Var}_I(\hat\sigma_y^{(j)})
 =L\cos^2(hI\tau).
\end{equation}
The corresponding local estimation uncertainty is
\begin{equation}
 \delta I_{\mathrm R,N}(I)
 =\frac{\sqrt{\operatorname{Var}_I(\hat M_{\mathrm R})}}
 {\sqrt{N}\,|\partial_I f_{\mathrm R}(I)|}
 =\frac{1}{h\tau\sqrt{NL}}.
 \label{eq:ramsey_uniform_uncertainty}
\end{equation}
Thus, for noninteracting spins under a uniform field, the local estimation uncertainty scales as $1/\sqrt{NL}$~\cite{huelga1997improvement}.

In practice, the spins occupy different positions near a current-carrying wire, and the magnetic field varies across the ensemble.
If the spatial extent of the ensemble is much smaller than its distance from the wire, the magnetic fields at different spin positions are nearly parallel. We therefore approximate their directions as common while retaining the position dependence of their magnitudes.
Let $\mathbf B_j(I)$ denote the magnetic field generated by the current $I$ at spin $j$. Since the field is proportional to the current, we write $\mathbf B_j(I)=I\mathbf b_j$, where $\mathbf b_j$ is the magnetic field per unit current at spin $j$. We then define the coupling strength as $h_j=\gamma_e|\mathbf b_j|$, where $\gamma_e>0$ is the electron gyromagnetic ratio.

The mean coupling strength is
\begin{equation}
 \bar h=\frac{1}{L}\sum_{j=1}^{L}h_j.
 \label{eq:ramsey_mean_coupling}
\end{equation}
Assuming the same magnetic-field direction for all spins while retaining the position-dependent coupling strengths, the independent-spin response is approximately
\begin{equation}
 f_{\mathrm R}(I)
 =\sum_{j=1}^{L}\sin(h_j I\tau).
 \label{eq:ramsey_nonuniform_response}
\end{equation}

When the coupling strengths are sufficiently close to $\bar h$, the response can be approximated as $f_{\mathrm R}(I)\simeq L\sin(\bar h I\tau)$, particularly near $I=0$.

\section{Quantum circuit learning}
\label{sec:qcl}

Quantum machine learning (QML) uses quantum systems to construct computational models for machine learning tasks~\cite{schuld2015introduction,schuld2019quantum,carleo2019machine,biamonte2017quantum,jerbi2023quantum}. In particular, quantum circuit learning (QCL) combines parameterized quantum circuits with classical optimization algorithms~\cite{mitarai2018quantum}. This quantum--classical optimization approach is an example of a variational quantum algorithm (VQA)~\cite{cerezo2021variational}. In QCL, a quantum circuit containing input-dependent and parameterized operations defines a learning model, and its parameters are optimized so that the model output approximates a prescribed target function. This framework can be applied to supervised learning, in which a set of known input--output pairs is used to train the quantum model.

In supervised learning, a training data set $\{(x_k,y_k)\}_{k=1}^{K}$ is given, where $x_k$ represents an input and $y_k$ its target output. A parameterized function $f_{\boldsymbol\theta}(x)$ is defined, and the parameters $\boldsymbol\theta$ are optimized so that $f_{\boldsymbol\theta}(x_k)$ approaches $y_k$. In QCL, the input data and trainable parameters are incorporated into quantum operations. Starting from an initial state $|0\cdots0\rangle$, a parameterized unitary operator $\hat U(x,\boldsymbol\theta)$ produces the output state. The learning model is defined as the expectation value of an observable $\hat M$:
\begin{equation}
f_{\boldsymbol\theta}(x)
=\langle0\cdots0|
\hat U^\dagger(x,\boldsymbol\theta)
\hat M
\hat U(x,\boldsymbol\theta)
|0\cdots0\rangle.
\label{eq:qcl_general_output}
\end{equation}

The agreement between the model output and the target values is evaluated using a cost function. A standard choice is the mean squared error,
\begin{equation}
\mathcal J(\boldsymbol\theta)
=\frac{1}{K}\sum_{k=1}^{K}
\left[f_{\boldsymbol\theta}(x_k)-y_k\right]^2.
\label{eq:qcl_general_loss}
\end{equation}
The parameters are updated using a classical optimization algorithm to minimize this cost function. After optimization, the trained quantum model can be used to produce outputs for new input values.

\section{Model and methods}
\label{sec:model_methods}

\begin{figure}[t]
 \centering
 \includegraphics[width=\columnwidth]{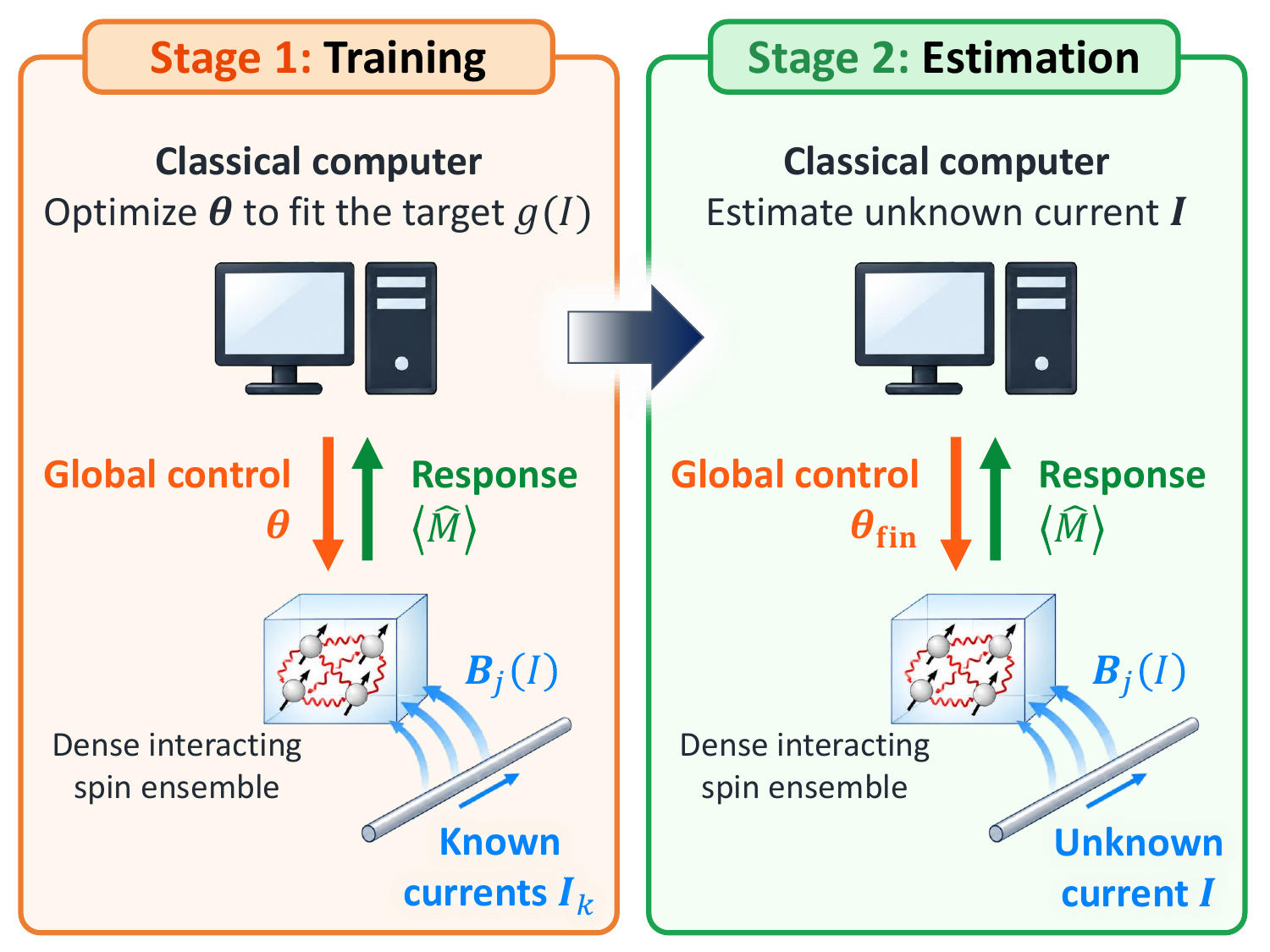}
 \caption{
    Schematic of the proposed quantum sensing method.
    A current $I$ in a nearby wire generates a position-dependent magnetic field $\boldsymbol B_j(I)$ acting on an interacting spin ensemble, where $j$ labels the spins and $\boldsymbol B_j(I)$ denotes the magnetic field acting on the $j$th spin.
    A parameterized global control field acts during sensing, and the collective magnetization $\hat M$ is measured at the end of the sensing interval.
    In Stage 1 (training), known currents $I_k$, with $k$ indexing the training samples, are used to optimize the global control parameters $\boldsymbol\theta$ so that the collective-magnetization response approaches the target function $g(I)$. In Stage 2 (estimation), the optimized parameters $\boldsymbol\theta_{\mathrm{fin}}$ are held fixed, and the response is used to estimate an unknown current $I$.
 }
 \label{fig.1}
\end{figure}

Our method consists of two stages: training and estimation, as illustrated in Fig.~\ref{fig.1}. During training, we use known current values to optimize the global control parameters so that the expectation value of the collective magnetization approaches a prescribed monotonic target function. After training, the optimized parameters are held fixed when estimating an unknown current.

\subsection{Physical model and sensing schemes}
\label{subsec:physical_model}

We consider $L$ spin-$1/2$ particles located near a straight current-carrying wire. The quantity to be estimated is the electric current $I$. The corresponding magnetic field acts locally on the spins, which also interact through magnetic dipole--dipole coupling. A spatially uniform, parameterized control field acts during sensing, and the collective $z$ magnetization is measured at the end.

Let $\tau$ denote the total sensing time and $t$ the time within the sensing interval ($0\leq t\leq \tau$). The Hamiltonian during sensing is then given by
\begin{equation}
 \hat H(I,\boldsymbol\theta;t)
 =\hat H_{\mathrm{in}}(I)
 +\hat H_{\mathrm{int}}
 +\hat H_{\mathrm{ctrl}}(\boldsymbol\theta;t).
 \label{eq:total_hamiltonian}
\end{equation}
We use angular-frequency units and set $\hbar=1$. The input and interaction Hamiltonians remain active during the global control operations.

Let \(\mathbf r_{\mathrm w}\) denote the position vector of a reference point on the straight wire, and let \(\hat{\mathbf u}_{\mathrm w}\) be the unit vector along the wire. We denote the position vector of spin \(j\) by \(\mathbf r_j\). The shortest displacement vector from the wire to spin \(j\) is then given by
\begin{equation}
 \boldsymbol\rho_j
 =\mathbf r_j-\mathbf r_{\mathrm w}
 -[(\mathbf r_j-\mathbf r_{\mathrm w})\cdot
   \hat{\mathbf u}_{\mathrm w}]\hat{\mathbf u}_{\mathrm w}.
\end{equation}
For an infinitely long straight wire, the Biot--Savart field at spin $j$ is
\begin{equation}
 \mathbf B_j(I)
 =\frac{\mu_0 I}{2\pi|\boldsymbol\rho_j|^2}
 \left(\hat{\mathbf u}_{\mathrm w}\times\boldsymbol\rho_j\right),
 \label{eq:biot_savart}
\end{equation}
where $\mu_0$ is the vacuum permeability. Writing $\hat{\boldsymbol\sigma}^{(j)}$ for the vector of Pauli operators on spin $j$, we have
\begin{equation}
 \hat H_{\mathrm{in}}(I)
 =\frac{\gamma_e}{2}
 \sum_{j=1}^{L}\mathbf B_j(I)\cdot
 \hat{\boldsymbol\sigma}^{(j)},
 \label{eq:input_hamiltonian}
\end{equation}
where $\gamma_e$ is the electron gyromagnetic ratio, with $\gamma_e/(2\pi)=28\,\mathrm{GHz\,T^{-1}}$.

The spin--spin interaction is modeled by
\begin{equation}
\begin{split}
 \hat H_{\mathrm{int}}=J_0\sum_{i<j}\frac{1}{r_{ij}^{3}}
 \bigl[&\hat{\boldsymbol\sigma}^{(i)}\cdot
        \hat{\boldsymbol\sigma}^{(j)}\\
 &-3(\hat{\boldsymbol\sigma}^{(i)}\cdot\hat{\mathbf r}_{ij})
    (\hat{\boldsymbol\sigma}^{(j)}\cdot\hat{\mathbf r}_{ij})\bigr],
\end{split}
 \label{eq:dipolar_hamiltonian}
\end{equation}
where $\mathbf r_{ij}=\mathbf r_i-\mathbf r_j$, $r_{ij}=|\mathbf r_{ij}|$, and $\hat{\mathbf r}_{ij}=\mathbf r_{ij}/r_{ij}$. We use the positive effective coefficient $J_0/(2\pi)=13\,\mathrm{MHz\,nm^3}$ in this Pauli-operator Hamiltonian. For $L=1$, the interaction term vanishes.

The global control consists of $D$ repetitions of $x$-, $y$-, and $z$-axis control fields, applied in this order. We denote the control parameters by $\boldsymbol\theta=\{\theta_{d,a}\}_{d=1,\ldots,D;\,a\in\{x,y,z\}}$, where $\theta_{d,a}$ is the control parameter for axis $a$ in repetition $d$. The sensing interval is divided into $3D$ equal slots of duration $\Delta t=\tau/(3D)$. During slot $(d,a)$, the control Hamiltonian is
\begin{equation}
 \hat H_{\mathrm{ctrl}}(\boldsymbol\theta;t)
 =-\frac{\theta_{d,a}}{2\Delta t}
 \sum_{j=1}^{L}\hat\sigma_a^{(j)}.
 \label{eq:control_hamiltonian}
\end{equation}
The same control field acts on all spins in each slot. Defining
\begin{equation}
\begin{split}
 \hat U_{d,a}(I,\boldsymbol\theta)
 =\exp\!\biggl\{-i\Delta t\biggl[
 &\hat H_{\mathrm{in}}(I)+\hat H_{\mathrm{int}}\\
 &-\frac{\theta_{d,a}}{2\Delta t}
  \sum_j\hat\sigma_a^{(j)}\biggr]\biggr\},
\end{split}
\end{equation}
the evolution operator is
\begin{equation}
 \hat U(I,\boldsymbol\theta)
 =\left(\hat U_{D,z}\hat U_{D,y}\hat U_{D,x}\right)
 \cdots\left(\hat U_{1,z}\hat U_{1,y}\hat U_{1,x}\right).
 \label{eq:time_evolution}
\end{equation}
Since the Hamiltonian is constant within each time slot, the total time evolution is exactly expressed as the ordered product of the corresponding evolution operators.

For the interacting sensor, the initial state is $|\psi_0\rangle=|0\rangle^{\otimes L}$. After time evolution, we measure the collective $z$ magnetization
\begin{equation}
 \hat M=\sum_{j=1}^{L}\hat\sigma_z^{(j)},
\end{equation}
whose expectation value defines the current response
\begin{equation}
 f_{\boldsymbol\theta}(I)
 =\langle\hat M\rangle_{\boldsymbol\theta,I}
 =\langle\psi_0|\hat U^\dagger(I,\boldsymbol\theta)
 \hat M\hat U(I,\boldsymbol\theta)|\psi_0\rangle.
 \label{eq:response}
\end{equation}
The corresponding variance of the measured observable is
\begin{equation}
 \operatorname{Var}_{\boldsymbol\theta,I}(\hat M)
 =\langle\hat M^2\rangle_{\boldsymbol\theta,I}
 -\langle\hat M\rangle_{\boldsymbol\theta,I}^{2}.
 \label{eq:magnetization_variance}
\end{equation}

We consider three control conditions for the interacting sensing protocol:
\begin{itemize}
 \item \emph{No control}: The interacting sensor with Hamiltonian $\hat H_{\mathrm{in}}(I)+\hat H_{\mathrm{int}}$ and no applied global control. This is the principal reference for assessing control-induced improvements.
 \item \emph{Initial}: The interacting sensor with randomly initialized control parameters $\boldsymbol\theta=\boldsymbol\theta_{\mathrm{init}}$.
 \item \emph{Final}: The interacting sensor with the final trained control parameters $\boldsymbol\theta=\boldsymbol\theta_{\mathrm{fin}}$, obtained as described in Sec.~\ref{subsec:learning_metrics}.
\end{itemize}

For these three sensing conditions, changing the sensing time $\tau$ can affect both the response slope and the magnetization variance, thereby altering the dynamic range and estimation uncertainty.

\subsection{Learning and performance evaluation}
\label{subsec:learning_metrics}

We construct a sinusoidal target using the characteristic current scale of the noninteracting Ramsey response introduced in Sec.~\ref{subsec:noninteracting_ramsey}. The parameter $A$ sets the target amplitude relative to the maximum collective magnetization $L$, while $\eta>0$ controls its width in current.

Using the approximation $f_{\mathrm R}(I)\simeq L\sin(\bar h I\tau)$ with $\bar h$ defined in Eq.~\eqref{eq:ramsey_mean_coupling}, we introduce the reference current $I_{\mathrm{ref}}=\pi/(2\bar h\tau)$, corresponding to the first maximum of the approximate Ramsey response. We then define $I_{\mathrm{tar}}=\eta I_{\mathrm{ref}}$. This idealized Ramsey model is used to set the training target and is not an additional sensing condition in the numerical comparisons. The target function is
\begin{equation}
 g(I)=AL\sin\!\left(\frac{\pi I}{2I_{\mathrm{tar}}}\right),
 \qquad -I_{\mathrm{tar}}\le I\le I_{\mathrm{tar}}.
 \label{eq:target}
\end{equation}
The target function increases monotonically from $-AL$ to $AL$ as $I$ varies from $-I_{\mathrm{tar}}$ to $I_{\mathrm{tar}}$.

Following the supervised-learning framework introduced in Sec.~\ref{sec:qcl}, we train $f_{\boldsymbol\theta}(I)$ using $K$ equally spaced known current values $I_k$ in this interval. The cost function combines a normalized fitting error with a variance penalty:
\begin{equation}
\begin{split}
 \mathcal J(\boldsymbol\theta)=
 &\frac{1}{K}\sum_{k=1}^{K}
 \left[\frac{f_{\boldsymbol\theta}(I_k)-g(I_k)}{L}\right]^2\\
 &+\frac{\lambda}{K}\sum_{k=1}^{K}
 \frac{\operatorname{Var}_{\boldsymbol\theta,I_k}(\hat M)}{L}.
\end{split}
 \label{eq:loss}
\end{equation}
The first term measures agreement with the target; the second penalizes collective-magnetization fluctuations, with strength $\lambda$.
Each component of the initial control parameters $\boldsymbol\theta_{\mathrm{init}}$ is sampled independently and uniformly from $[-\pi,\pi)$.
We optimize the control parameters $\boldsymbol\theta$ by minimizing $\mathcal J(\boldsymbol\theta)$ and obtain the final parameters $\boldsymbol\theta_{\mathrm{fin}}$. These parameters are then held fixed when estimating an unknown current.

We evaluate sensing performance using two quantities: the dynamic range (DR) and the DR-averaged estimation uncertainty $\overline{\delta I}_{\mathrm{DR}}$.

To determine these quantities, we first calculate the per-measurement local estimation uncertainty $\delta I(I)\equiv\delta I_{N=1}(I)$ by setting $N=1$ in Eq.~\eqref{eq:metrology_local_uncertainty}, rather than using a time-normalized sensitivity.
In a practical experiment, each measurement cycle would also require state preparation and readout, giving a total duration $t_{\mathrm{cycle}}=t_{\mathrm{prep}}+\tau+t_{\mathrm{read}}$.
If $t_{\mathrm{prep}}$ and $t_{\mathrm{read}}$ are much larger than $\tau$, $t_{\mathrm{cycle}}$ is almost independent of $\tau$.
In this case, for a fixed total experimental time $T_{\mathrm{total}}$, the number of repetitions $N$ is approximately $T_{\mathrm{total}}/(t_{\mathrm{prep}}+t_{\mathrm{read}})$, which is almost constant across different sensing times.
Under this assumption, we neglect differences in the number of measurement repetitions across the sensing times considered in our performance comparison.

We define a monotonic branch as a connected interval of input current over which the sensor response is strictly monotonic.
For DR evaluation, we generally consider the monotonic branch containing $I=0$.
Within the branch under consideration, we determine the minimum local estimation uncertainty $\delta I_{\min}$, and identify connected subintervals satisfying

\begin{equation}
    \delta I(I)\le 2\delta I_{\min}.
\end{equation}
For the branch containing $I=0$, we select the widest qualifying subinterval, which need not contain $I=0$ itself.
For a selected subinterval $[I_-,I_+]$, the dynamic range (DR) is defined as
\begin{equation}
    \mathrm{DR}=I_+-I_-.
\end{equation}
The selection procedure for the turning-point case and the numerical details are described in Sec.~\ref{subsec:numerical_settings}.

We define the mean of the local estimation uncertainty over the selected DR interval as
\begin{equation}
 \overline{\delta I}_{\mathrm{DR}}
 =\frac{1}{\mathrm{DR}}\int_{I_-}^{I_+}\delta I(I)\,dI.
 \label{eq:average_delta_i}
\end{equation}
Hereafter, we use the DR-averaged quantity $\overline{\delta I}_{\mathrm{DR}}$ as the estimation uncertainty in our performance comparisons.

\subsection{Numerical settings}
\label{subsec:numerical_settings}

Spin configurations are generated by independently sampling the position of each spin from an isotropic three-dimensional Gaussian distribution centered at the origin, with a standard deviation of $3\,\mathrm{nm}$. Configurations with any spin separation below $0.15\,\mathrm{nm}$ are rejected. The straight wire runs parallel to the $y$-axis and passes through $(0,0,30)\,\mathrm{nm}$.
For each number of spins $L$, we generate 20 independently sampled spin configurations.
For the optimization runs with $L=2,3,4,5$, each realization also includes a randomly initialized set of control parameters.
Within each realization, the same spin configuration is used for the No control, Initial, and Final conditions.
The No control condition depends only on the spin configuration.
For a given $L$, the same realizations are used for comparisons across different sensing times and variance-penalty strengths.

For the primary optimization results, we use $L=2,3,4,5$, $D=20$, $\tau=1\,\mu\mathrm{s}$, $K=30$, $A=1$, and $\eta=8$.
We take $\lambda\in\{0,0.01,0.1\}$.
All numerical simulations were implemented in Python 3. The control parameters were optimized using the derivative-free COBYLA algorithm~\cite{powell1994cobyla}, as implemented in SciPy 1.17.0~\cite{virtanen2020scipy} based on PRIMA~\cite{Zhang_2023}, with at most 1000 iterations.

The response, variance, and local estimation uncertainty $\delta I(I)$ are evaluated on an equally spaced current grid over $I\in[-60,60]\,\mu\mathrm A$, with a spacing of $0.05\,\mu\mathrm A$. These values are used to determine the monotonic branches and DR intervals, and to calculate the estimation uncertainty $\overline{\delta I}_{\mathrm{DR}}$ used in the performance comparisons.
The integral in Eq.~\eqref{eq:average_delta_i} is evaluated using the trapezoidal rule, with interpolation at the DR boundaries. Statistical summaries are reported as medians and interquartile ranges (IQRs).

To examine the dependence of DR on the number of spins in the No control condition, we use $L\in\{1,2,3,4,5\}$.
For $L=1$, the response is even in $I$ and has a turning point at $I=0$, consistent with the single-spin behavior discussed in Sec.~\ref{subsec:single_spin_cosine}. We therefore evaluate its positive- and negative-current monotonic branches separately and select the qualifying interval closest to $I=0$. If the distances are equal, we select the negative-current side as a tie-breaking convention.

For the sensing-time comparison, we use $L=4$. The Final condition is evaluated at $\tau\in\{1,2,4\}\,\mu\mathrm s$ for each of the three loss weights, while the No control condition is evaluated at $\tau\in\{1/10,1/8,1/6,1/4,1/2,1\}\,\mu\mathrm s$.

\section{Numerical results and discussion}
\label{sec:numerical_results}

\begin{figure}[t]
 \centering
 \includegraphics[width=\columnwidth]{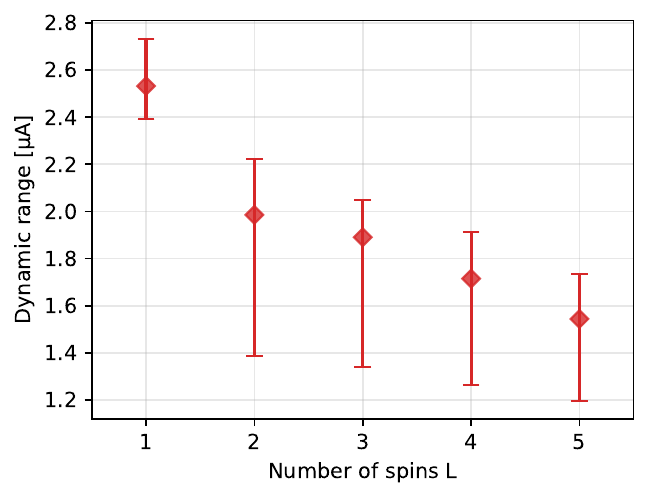}
 \caption{Dependence of DR on the number of spins $L$ for the No control condition at $\tau=1\,\mu\mathrm{s}$. Markers indicate the median over 20 independently sampled spin configurations at each $L$; vertical error bars show the interquartile range (IQR).}
 \label{fig:no-control-dr}
\end{figure}

Before applying global control, we characterize the DR of the sensor in the No control condition.
Figure~\ref{fig:no-control-dr} shows its dependence on the number of spins $L$.
The median DRs for $L=1,2,3,4,5$ are approximately $2.53$, $1.99$, $1.89$, $1.71$, and $1.54\,\mu\mathrm{A}$, respectively.
The median DR decreases monotonically with increasing $L$ over the range studied.
Since changing $L$ also changes the spin configurations and their couplings to the input field, this comparison does not isolate the effect of dipolar interactions.
Nevertheless, all multi-spin cases studied here show smaller median DRs than the single-spin case under the No control condition.

\begin{figure*}[t]
 \centering
 \begin{tabular}{@{}cc@{}}
  \makebox[0.48\textwidth][l]{\textbf{(a)}} &
  \makebox[0.48\textwidth][l]{\textbf{(b)}} \\
  \includegraphics[width=0.48\textwidth]{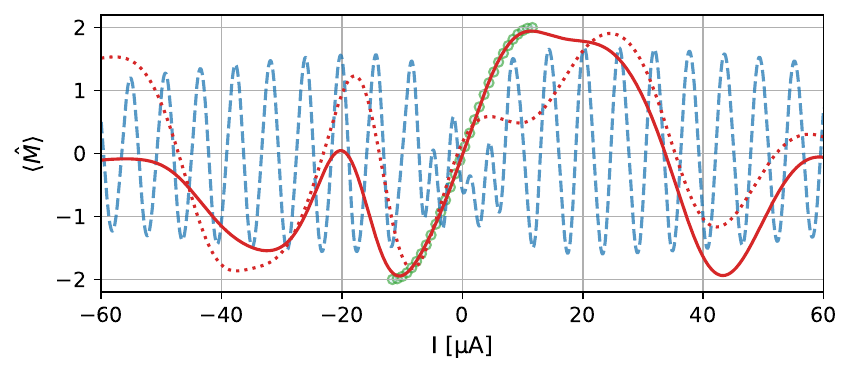} &
  \includegraphics[width=0.48\textwidth]{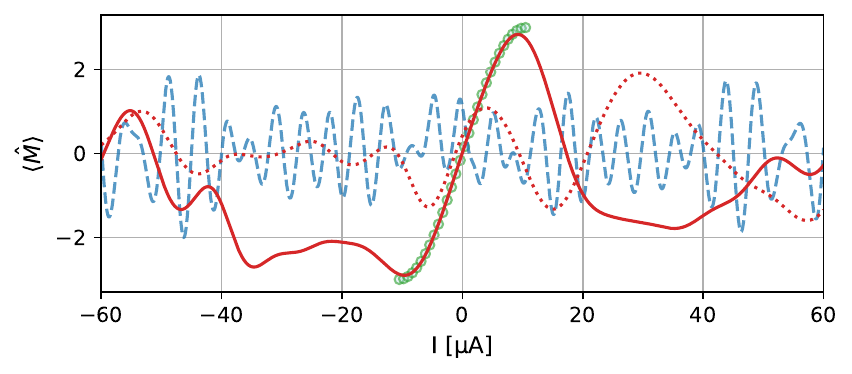} \\
  \makebox[0.48\textwidth][l]{\textbf{(c)}} &
  \makebox[0.48\textwidth][l]{\textbf{(d)}} \\
  \includegraphics[width=0.48\textwidth]{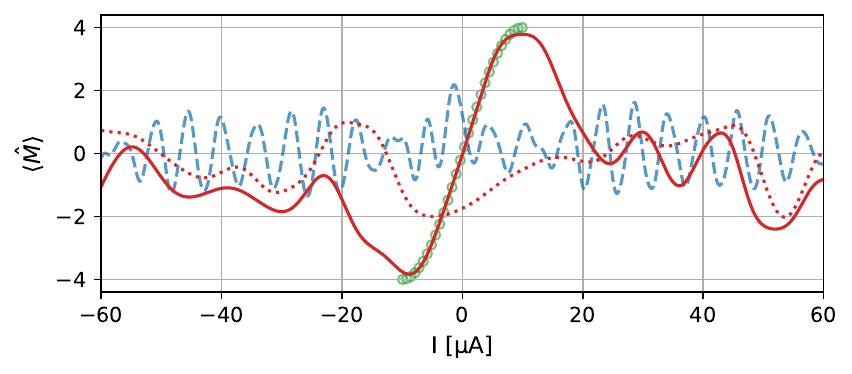} &
  \includegraphics[width=0.48\textwidth]{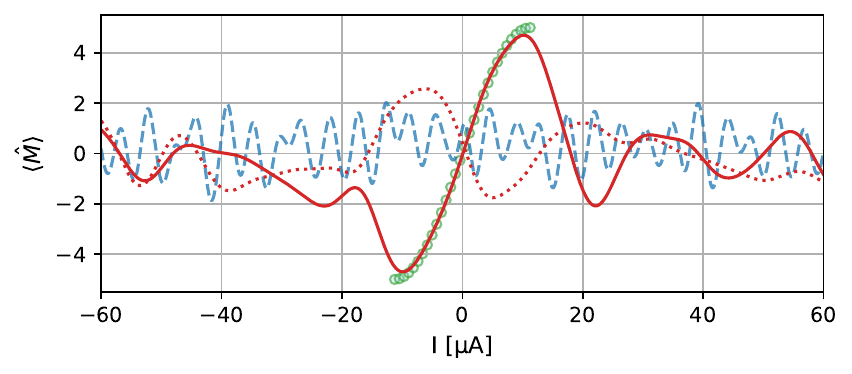}
 \end{tabular}
 \par\vspace{2mm}
 \includegraphics[width=0.50\textwidth]{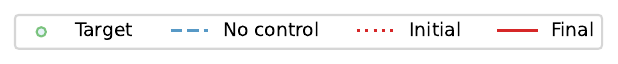}
 \caption{Current responses for (a) $L=2$, (b) $L=3$, (c) $L=4$, and (d) $L=5$. The target response $g(I)$ is shown together with the responses for the No control, Initial, and Final conditions. The optimization uses $\tau=1\,\mu\mathrm{s}$ and $\lambda=0$. Each panel shows one representative realization whose final training cost is near the median for that number of spins.}
 \label{fig:response-shaping}
\end{figure*}

We next investigate how global control changes the current response.
Figure~\ref{fig:response-shaping} compares the No control, Initial, and Final conditions with the target response for $L=2,3,4,5$.
The No control curves show repeated oscillations with current.
In the Initial condition, global control changes the response, but the resulting curves do not closely follow the target.
After optimization, the Final response follows the monotonic target over the training interval in the representative realization shown for each $L$.
Comparing Initial with Final therefore identifies the change produced by optimization, rather than by the mere presence of a control field.
We next examine whether these changes also improve DR and estimation uncertainty across the sampled realizations.

\begin{figure*}[t]
 \centering
 \begin{tabular}{@{}cc@{}}
  \makebox[0.48\textwidth][l]{\textbf{(a)}} &
  \makebox[0.48\textwidth][l]{\textbf{(b)}} \\
  \includegraphics[width=0.48\textwidth]{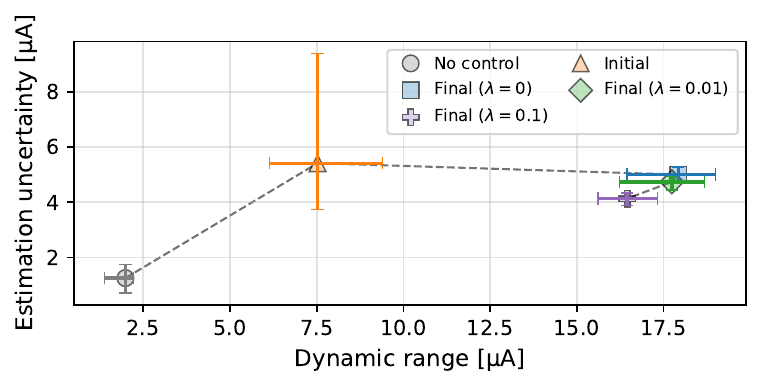} &
  \includegraphics[width=0.48\textwidth]{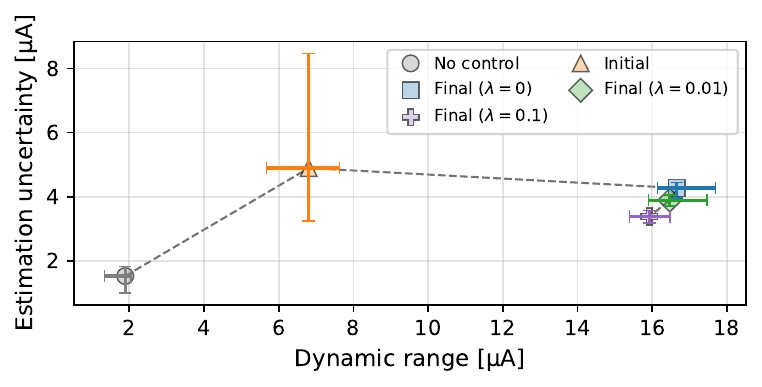} \\
  \makebox[0.48\textwidth][l]{\textbf{(c)}} &
  \makebox[0.48\textwidth][l]{\textbf{(d)}} \\
  \includegraphics[width=0.48\textwidth]{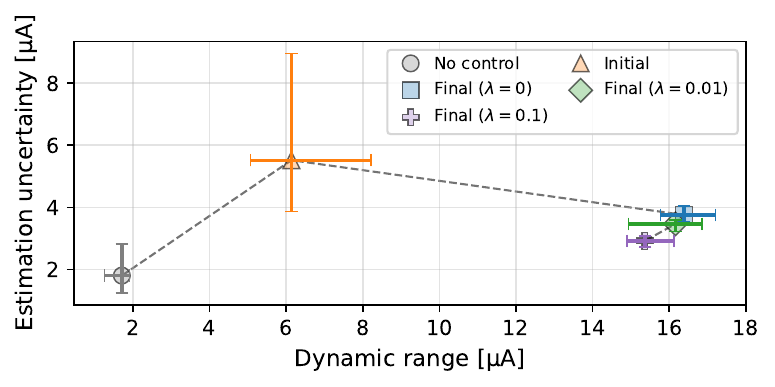} &
  \includegraphics[width=0.48\textwidth]{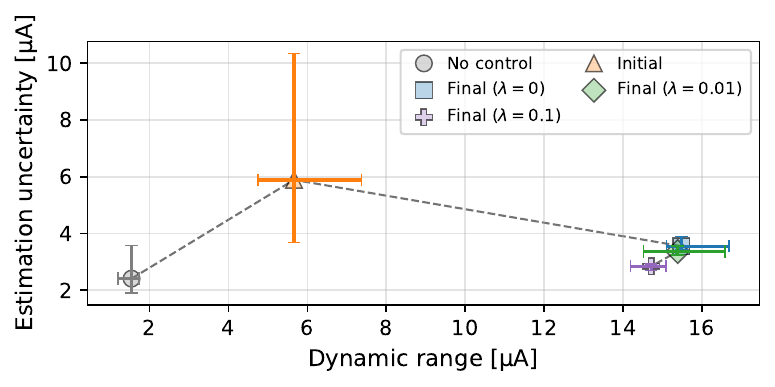}
 \end{tabular}
\caption{DR and estimation uncertainty for (a) $L=2$, (b) $L=3$, (c) $L=4$, and (d) $L=5$ at $\tau=1\,\mu\mathrm{s}$. The No control and Initial conditions are compared with Final at $\lambda\in\{0,0.01,0.1\}$. Each marker shows the median over 20 realizations; horizontal and vertical error bars show the interquartile ranges.
The horizontal axis shows the dynamic range (DR), and the vertical axis shows the estimation uncertainty $\overline{\delta I}_{\mathrm{DR}}$. Dashed lines connect selected median points to aid visual comparison.}
 \label{fig:variance-tradeoff}
\end{figure*}

To evaluate the sensing performance, we compare DR with the estimation uncertainty $\overline{\delta I}_{\mathrm{DR}}$ defined in Eq.~\eqref{eq:average_delta_i}.
Figure~\ref{fig:variance-tradeoff} shows these quantities for the No control, Initial, and Final conditions.
For all four numbers of spins, Initial has a larger median DR but a larger median estimation uncertainty than No control.
For $\lambda=0$, Final has a larger median DR and lower median estimation uncertainty than Initial.
Thus, the changes induced by random control and those produced by optimizing its parameters have different effects on sensing performance.

We next examine how the variance penalty affects DR and estimation uncertainty.
For each $L$, Final with $\lambda=0.1$ has a lower median estimation uncertainty than Final with $\lambda=0$, but its median DR is also smaller.
Since the local estimation uncertainty depends on both the magnetization variance and the response slope, as shown in Eq.~\eqref{eq:metrology_local_uncertainty}, reducing the variance alone does not guarantee an improvement. Nevertheless, the results show that adding a variance penalty to the response-fitting cost can reduce the estimation uncertainty, albeit at the expense of DR.

\begin{figure*}[t]
 \centering
 \includegraphics[width=0.8\textwidth]{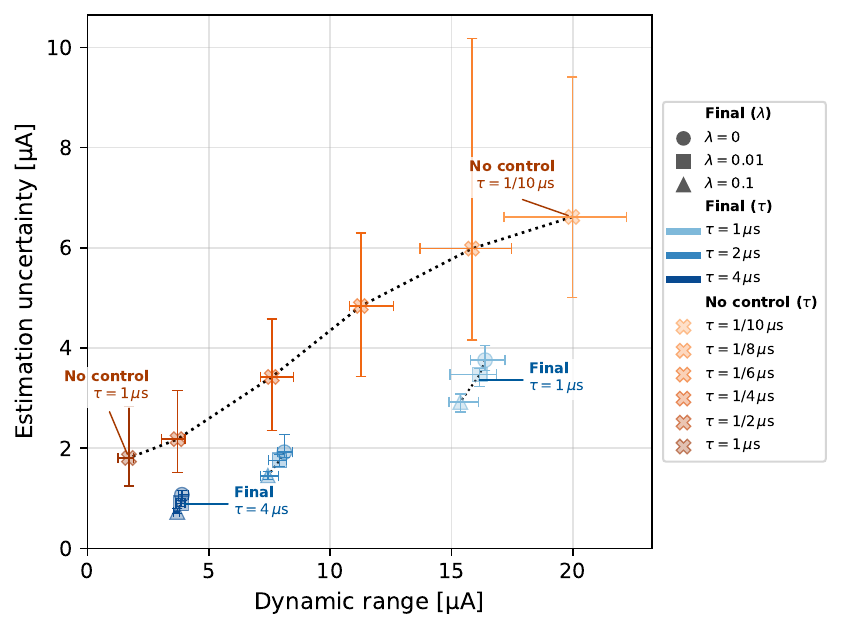}
 \caption{DR and estimation uncertainty at different sensing times for $L=4$. The Final condition is evaluated at $\tau\in\{1,2,4\}\,\mu\mathrm{s}$ for $\lambda\in\{0,0.01,0.1\}$. The No control condition is evaluated at $\tau\in\{1/10,1/8,1/6,1/4,1/2,1\}\,\mu\mathrm{s}$. Each marker represents the median of 20 realizations; horizontal and vertical error bars indicate the interquartile ranges. The horizontal axis shows the dynamic range (DR), and the vertical axis shows the estimation uncertainty $\overline{\delta I}_{\mathrm{DR}}$. The dotted line connects the No control medians in order of sensing time.}
 \label{fig:time-comparison}
\end{figure*}

We finally compare optimized global control with a simpler way of extending DR: shortening the sensing time $\tau$ in the No control condition.
Figure~\ref{fig:time-comparison} shows the resulting DR and estimation uncertainty for $L=4$.
For both conditions, shorter sensing times yield larger median DRs but also larger median estimation uncertainties.
For DRs between approximately $3$ and $16\,\mu\mathrm{A}$, suitable choices of $\tau$ allow the Final condition to achieve both a larger DR and a smaller median estimation uncertainty than the No control condition.
Furthermore, for the Final condition at $\tau=4\,\mu\mathrm{s}$, increasing $\lambda$ reduces the median estimation uncertainty with a relatively small reduction in DR.
These results indicate that optimized global control provides an additional means of improving the DR--uncertainty trade-off under the conditions studied here.

\FloatBarrier
\section{Conclusion}
\label{relabel}
We have proposed a quantum-circuit-learning-inspired method to improve the dynamic range of interacting quantum sensors. In dense spin ensembles, dipole--dipole interactions can induce complicated oscillations in the sensor response, limiting the range over which an unknown external field can be estimated unambiguously. To overcome this limitation, we optimize parameterized global control fields applied during the sensing process so that the expectation value of the collective magnetization approximates a prescribed monotonic function of the input field. Importantly, our approach does not require individual control of the sensing spins or the suppression of their mutual interactions.

Through numerical simulations of dipolar-interacting spin ensembles, we demonstrated that the optimized global control can substantially reshape the sensor response and extend its dynamic range. By comparing the optimized control with randomly initialized control, we confirmed that the optimization itself improves the trade-off between dynamic range and estimation uncertainty. We also investigated the effect of introducing a variance penalty into the optimization cost function. Increasing this penalty reduces the estimation uncertainty but generally decreases the dynamic range, indicating that these two performance measures must be optimized jointly. Furthermore, comparison with an uncontrolled sensor operated at shorter sensing times revealed that our method can achieve a lower estimation uncertainty at a comparable dynamic range under the conditions studied.

These results demonstrate that learning-based optimization of the sensor response provides a promising strategy for mitigating dynamic-range limitations in interacting quantum sensors.
Our approach offers a complementary means of improving quantum sensing performance through global quantum control, potentially enabling the use of densely packed spin ensembles for high-spatial-resolution sensing over an extended range of external fields.

\begin{acknowledgments}
Hideaki Kawaguchi, Takahiko Satoh, and Yuichiro Matsuzaki are supported by JST Moonshot R\&D Grant Number JPMJMS226C. 
Hideaki Kawaguchi is also supported by JSPS KAKENHI Grant Number 23K16996. Hideaki Kawaguchi is also supported by the Center of Innovation for Sustainable Quantum AI (JST Grant Number JPMJPF2221). Takahiko Satoh is also supported by JST COI-NEXT Grant Number JPMJPF2221. Yuichiro Matsuzaki is also supported by JSPS KAKENHI Grant Number 23H04390, JST CREST Grant Number JPMJCR23I5, and Presto JST Grant Number JPMJPR245B.


\end{acknowledgments}

\bibliography{ref}

\end{document}